\documentclass[10pt,prl,aps,twocolumn,showpacs,floatfix]{revtex4}
\usepackage{graphicx}
\usepackage{amssymb}
\usepackage{amsmath}

\begin{document}

\title{Constrained sampling method for analytic continuation}

\author{Anders W. Sandvik}
\affiliation{Department of Physics, Boston University, 590 Commonwealth Avenue, Boston, Massachusetts 02215}
\affiliation{Institute of Physics, Chinese Academy of Sciences, P.O. Box 603, Beijing 100190, China}

\begin{abstract}
A method for analytic continuation of imaginary-time correlation functions (here obtained in quantum Monte Carlo
simulations) to real-frequency spectral functions is proposed. Stochastically sampling a spectrum parametrized by a large 
number of delta-functions, treated as a statistical-mechanics problem, it avoids distortions caused by (as demonstrated here) 
configurational entropy in previous sampling methods. The key development is the suppression of entropy by constraining the
spectral weight to within identifiable optimal bounds and imposing a set number of peaks. As a test case, the dynamic 
structure factor of the $S=1/2$ Heisenberg chain is computed. Very good agreement is found with Bethe Ansatz results 
in the ground state (including a sharp edge) and with exact diagonalization of small systems at elevated temperatures.
\end{abstract}

\date{\today}


\maketitle

Obtaining real-frequency dynamic response functions from imaginary-time correlations remains one of the outstanding challenges 
for quantum Monte Carlo (QMC) and related simulation methods (e.g., lattice QCD). The general form of the problem is to invert 
the relationship
\begin{equation}
G(\tau)=\int d\omega A(w)K(\tau,\omega),
\label{garel}
\end{equation}
where a QMC estimate $\tilde G(\tau)$ of the correlation function $G(\tau)$ is available, $A(\omega)$ is the spectral function 
sought, and the kernel $K(\tau,\omega)$ depends on the type of spectral function. Similar to an inverse Laplace transform,
there is no closed form for $A(\omega)$. Only broad features of $A(\omega)$ can be resolved in numerical analytic continuation,
because information on fine structure is only present at a level of precision of $G(\tau)$ which is not attainable in practice. 
Nevertheless, one can extract important dynamical features and the key question is how to do that with the maximum fidelity, given 
$\tilde G(\tau)$ and its statistical errors. Significant progress will be presented here.

The Maximum Entropy (ME) method \cite{gull1984} was adapted to the particulars of QMC some time ago \cite{silver90}. Overcoming problems 
of previous approaches \cite{schuttler85,white89}, it quickly became a standard tool \cite{jarrell96}. The ME method has an appealing footing 
in probability theory, but in many cases the entropic prior regularizes the spectrum too heavily, leading to excessive broadening and 
distortions. To avoid this, an alternative line of methods has been developed \cite{sandvik98,beach04,syljuasen08,fuchs10,wu13} (and applied to 
diverse systems \cite{reichman09,aristov10,feldner11,goth13}) which do not impose the entropic prior, instead using stochastic sampling of 
$A(\omega)$ with the probability distribution
\begin{equation}
P(A) \propto {\rm exp}(-\chi^2/2\Theta),
\label{psample1}
\end{equation}
where $\chi^2$ is the standard measure of the goodness of the fit of $G(\tau)$ obtained from $A(\omega)$ according to Eq.~(\ref{garel}) 
to the QMC-computed $\tilde G(\tau)$ with its full covariance matrix \cite{jarrell96,syljuasen08} for a set $\{\tau_i\}$. The spectrum
is typically parametrized as a sum of a large number of $\delta$-functions, though other forms have also been used \cite{wu13}. 
The sampling temperature $\Theta$ in Eq.~(\ref{psample1}) acts as a regularizing parameter. 

An important insight was gained by Beach \cite{beach04}, showing that a mean-field treatment of the sampling approach gives the ME method, with 
$\Theta$ corresponding to the entropic weight. Subsequently, Sylju{\aa}sen argued for fixing $\Theta=1$ \cite{syljuasen08} (as had also been 
done by White in earlier work \cite{white91}). A recent variant of the method by Fuchs {\it et al.}~uses 
Bayesian inference to determine $\Theta$ \cite{fuchs10}.

Here a previously overlooked problem with the sampling approach is pointed out, and a solution is offered which improves the performance 
to the point that s sharp edge of the spectrum can be resolved without imposing it \cite{sandvik01} by some functional form. The key 
insight is that, when parametrizing $A(\omega)$ with $N$ $\delta$-functions and treating these as the configuration space of a statistical 
mechanics problem with $\chi^2$ corresponding to the energy, the configurational entropy (not to be confused with the information entropy 
of the ME method) increases when $N$ is increasing, thereby forcing $A(\omega)$ away from a good fit. This happens primarily 
because $\chi^2$ does not have the normal extensive property of an energy function. Spectral weight is therefore forced out by entropic 
pressure beyond the bounds of the true spectrum, leading also to severe distortions of other parts of the spectrum. Ways to counteract
this entropic catastrophe will be presented.

{\it Model and method.}---The method will here be demonstrated for the dynamic spin structure factor of the $S=1/2$ 
Heisenberg spin chain, with Hamiltonian
\begin{equation}
H = \sum_{i=1}^L {\bf S}_i \cdot {\bf S}_j.
\end{equation}
The stochastic series expansion QMC algorithm \cite{sse} is used to compute the correlation function
\begin{equation}
G_q(\tau) = \langle S^z_{-q}(\tau)S^z_{q}(0)\rangle,
\end{equation}
where $S^z_q$ is the Fourier transform of the spins. With the kernel $K(\tau,\omega) = \pi^{-1}{\rm e}^{-\tau\omega}$ in Eq.~(\ref{garel}) 
and $\omega \in (-\infty,\infty)$, $A(\omega)$ is the dynamic structure factor $S(q,\omega)$. At inverse temperature $\beta=1/T$ it 
satisfies $S(q,-\omega) = {\rm e}^{-\beta \omega}S(q,\omega)$. In the method to be discussed, it is more practical to define 
$A_q(\omega) = S(q,\omega)(1+{\rm e}^{-\beta \omega})$, so that
\begin{equation}
K(\tau,\omega) = ({\rm e}^{-\tau\omega} + {\rm e}^{-(\beta-\tau)\omega})(1 + {\rm e}^{-\beta\omega})^{-1} \pi^{-1},
\end{equation}
and integrating over $\omega \in (0,\infty)$ in Eq.~(\ref{garel}).

$G_q(\tau)$ is computed for a set $\tau \in \{\tau_1,\ldots,\tau_{M}\}$ with $\tau_j = (j-1)\Delta_\tau$, and, because of symmetry
properties, only the range $0 \le \tau \le \beta/2$ has to be considered. For large $\tau$ the statistical errors may become too large, and he number 
of points $M$ is therefore adjusted in this work so that the relative error never exceeds $10\%$. 

With $A_q(\omega)$ parametrized as
\begin{equation}
A_q(\omega)=\sum_{n=1}^N a_n \delta(\omega - \omega_n),~~~\omega_n = (n-1/2)\Delta_\omega,
\label{aqform}
\end{equation}
the weights $\{ a_n\}$ will first be importance-sampled using Eq.~(\ref{psample1}) with $\Theta=1$ and later with a modified form.
Different types of updates are carried out to transfer weight between two or more $\delta$-functions, with the normalization $G_q(0)$ 
conserved to achieve a high acceptance rate \cite{sandvik98,syljuasen08}. Conservation of higher moments can also be incorporated \cite{sandvik98} 
but will not be done here. Single-weight updates account for the (small) normalization fluctuations.

$T=0$ results for $S(q,\omega)$ are available from Bethe Ansatz (BA) calculations including two-and four-spinon processes, which accounts
for almost all spectral weight \cite{caux05}. Comparisons will be made with these results for a system with $500$ spins \cite{cauxdata}
as well as with exact diagonalization results for an $L=16$ chain at $T>0$ \cite{starykh97}.

\begin{figure}
\centerline{\includegraphics[width=7cm, clip]{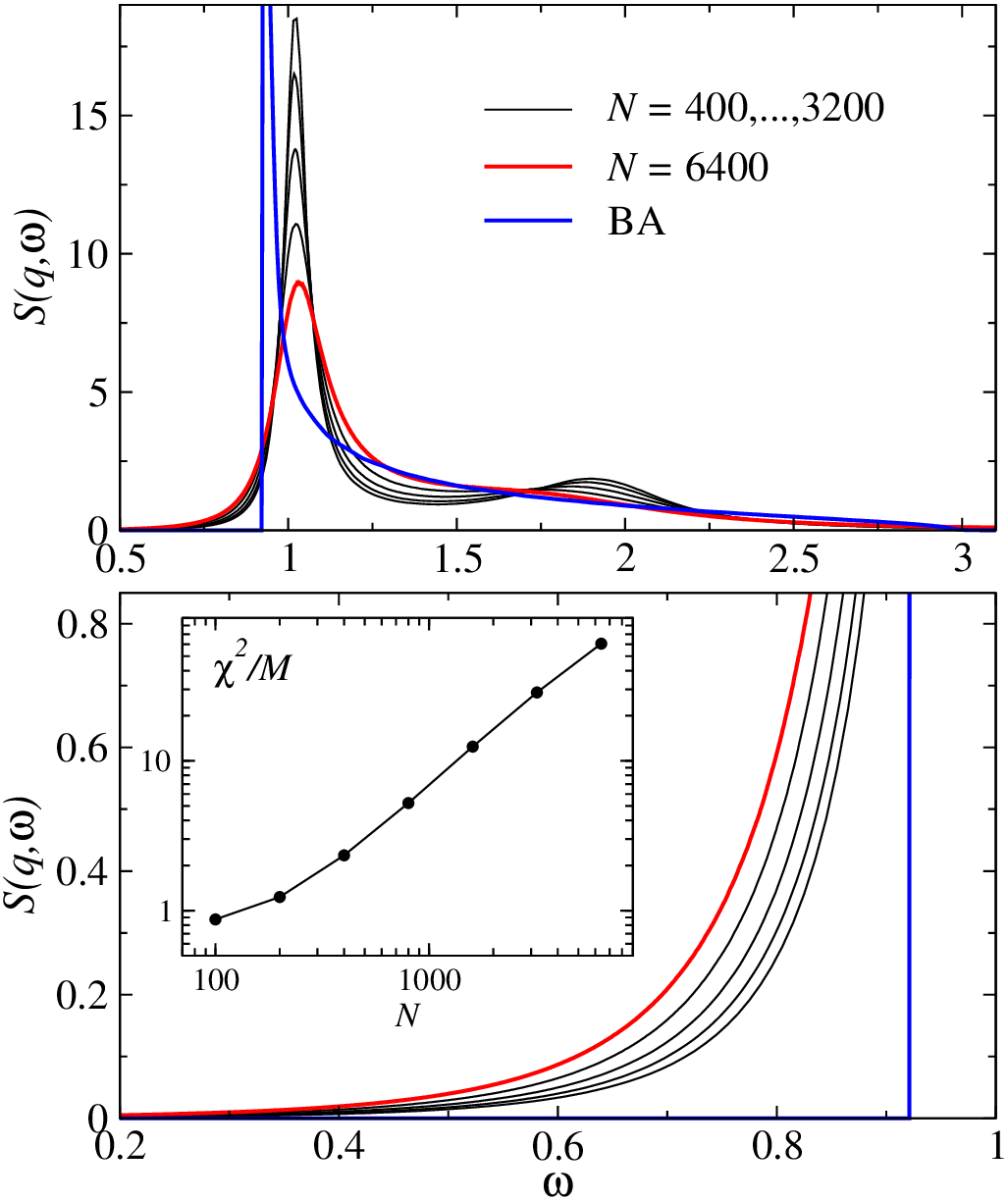}}
\vskip-1mm
\caption{(Color online) Dynamic structure factor at $q=0.8\pi$ obtained by unconstrained sampling for$\omega \in [0,4]$ 
and different $N$ of the form $100\times 2^n$ (peak decreasing with increasing $N$), compared with a BA 
result \cite{caux05,cauxdata}. The lower panel shows details of the low-frequency part. The inset shows the goodness of the fit versus $N$.}
\label{fig1}
\vskip-2mm
\end{figure}

{\it Unconstrained sampling}.---To illustrate the entropic problem with the sampling method in the $\Theta=1$ formulation \cite{syljuasen08}, 
results for $L=500$, $q=0.8\pi$ are shown in Fig.~\ref{fig1}. The QMC calculations were carried out at inverse temperature $\beta=500$, which 
for all practical purposes gives $T=0$ results for $G_q(\tau)$ at the momentum considered. The time spacing was $\Delta_\tau=1/4$ and the 
number of data points $M=33$. The relative statistical error of $G_q(\tau)$ was $\approx 10^{-5}$ at $\tau_1=0$ and $\approx 0.1$ at $\tau_M$. 
Fig.~\ref{fig1} shows results obtained with several different numbers of $\delta$-functions in the spectrum. Comparing with the BA result, 
a striking feature is how the low-energy weight in the region below the actual 
spectral edge increases with increasing $N$ (and the weight similarly increases also above the upper bound at $\omega \approx 3$), while 
the peak is suppressed. The main peak is too far to the right, and there is a second, spurious peak at higher $\omega$ which is more 
prominent for small $N$. Overall, the results look similar to those of Ref.~\cite{syljuasen08}, where only a fixed $N=1000$ was used.

From a statistical-mechanics point of view, it is clear that the sampling method suffers an entropic catastrophe for large $N$, 
with growing weight outside the bounds of the actual spectrum and, therefore, a rapidly increasing $\chi^2$. Results indicating a similar
problem with the Bayesian selection of $\Theta$ can be seen in Fig.~7 of Ref.~\cite{fuchs10}. To counteract the entropy, 
several modifications of the sampling method will be introduced next.

{\it Constrained sampling at T=0}.---If the spectral bounds are known one can prevent the entropy-driven leakage of weight and, 
presumably, the associated distortions of the spectrum within the bounds. Normally the bounds are not known, 
however, but, as will be shown below, they can be approximately determined using the data. Before discussing how this is done, another 
important feature reducing the configurational entropy will be incorporated.

With the spectrum parametrized as in (\ref{aqform}), no particular shape is imposed and when $N$ becomes sufficiently
large any spectrum can be reproduced in principle. In practice, however, one can only hope to resolve some prominent features of the
spectrum. In particular, it is difficult to resolve a large number of closely spaced peaks. In many cases one has some prior 
information, e.g., one may know that the spectrum should have one or two peaks. In other cases, recognizing the generic limitations of
analytic continuation, one may want to use a spectrum with the smallest number of peaks consistent with the QMC data. It is
easy to impose a fixed number of peaks in sampling a $\delta$-function sum (\ref{aqform}), by starting with a spectrum with the 
desired number of peaks and only proposing updates which do not create or destroy peaks. Here a one-peak spectrum $A_q(w)$
will be considered [which implies a single peak also in $S(q,\omega)$, unless $T$ is very high and a small peak at low $\omega$
can appear], but the procedures can be very easily generalized to any number of peaks. 

\begin{figure}
\centerline{\includegraphics[width=7.25cm, clip]{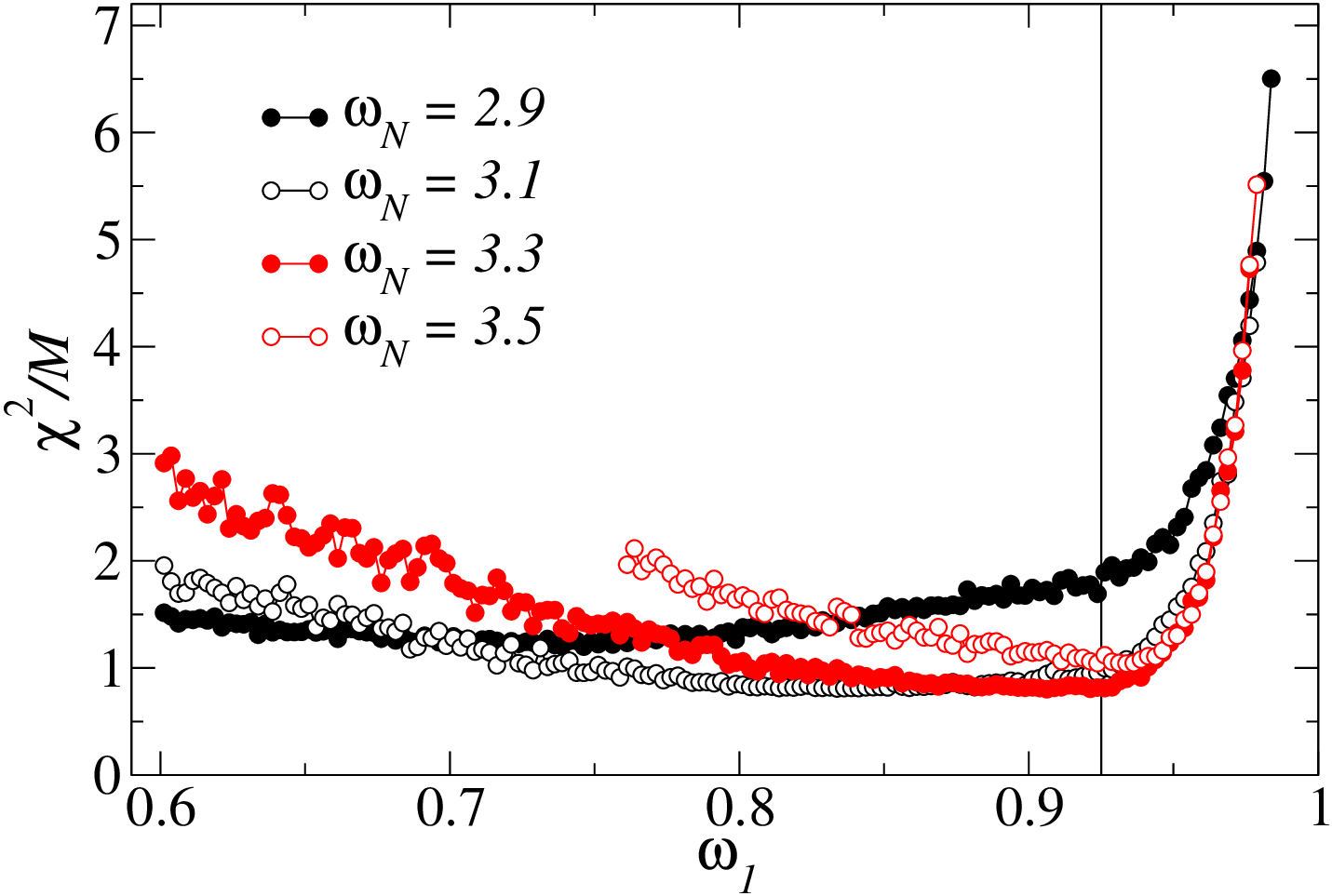}}
\vskip-1mm
\caption{(Color online) Goodness of fit versus the lower bound of the spectrum for an $L=500$ chain
at $q=0.8\pi$, for several choices of the upper bound $\omega_N$ and $\Delta_\omega=0.0025$. The vertical
line shows the location of the edge of the BA spectrum.}
\label{fig2}
\vskip-2mm
\end{figure}

The bounds of the spectrum can be approximately determined by following the goodness of the 
fit as a function of the frequencies $\omega_1$ and $\omega_N$ in Eq.~(\ref{aqform}). Fixing one of the bounds, $\omega_N$ say, a
minimum in $\chi^2$ versus $\omega_1$ has to exist for large $N$, because the entropic effect is reduced as $\omega_1$ is increased (provided of 
course that the true spectrum has vanishing or very small low-frequency weight), thereby reducing $\chi^2$ until $\omega_1$ starts to extend 
into the region of significant weight, whence $\chi^2$ must increase. Fig.~\ref{fig2} shows results of such scans for the normalized goodness 
of fit, $\chi^2/M$ (with $M$ used instead of the unknown number of degrees of freedom, $N_{\rm dof}$ \cite{jarrell96}). The minimum $\chi^2/M$ 
is indeed for $\omega_1$ close to the lower spectral edge, and there is a sharp increase when $\omega_1$ is pushed beyond the edge. The upper
edge can be roughly determined to within $5-10\%$ of the location of the sharp decay in weight at $\omega\approx 3.0$ in the BA spectrum.
The $\chi^2$ minimum becomes more prominent for large $N$ (hence making it easier to determine the bounds), in accord with the entropic scenario.

When determining the spectral bounds it is safe to allow $\chi^2$ to deviate by a statistically insignificant amount 
$\propto M^{1/2}$ from the best value $\chi^2_{\rm min}$ [given that the width of the $\chi^2$ distribution is $(2N_{\rm dof})^{1/2}$ and $M \sim N_{\rm dof}$], 
going toward higher $\omega_1$ where $\chi^2$ grows very rapidly, and also toward higher $\omega_N$ where the spectrum is less sensitive to the exact 
location of the bound. For the lower bound in the case of a spectrum with a sharp edge, as is the case here, one should not push $\omega_1$ beyond the 
point where the peak of the spectrum is at the lower bound. One may also determine $\omega_1$ by separately analyzing the large-$\tau$ behavior, though 
that is not always an easy task unless the lower edge is a well isolated $\delta$-function.

\begin{figure}
\centerline{\includegraphics[width=7.25cm, clip]{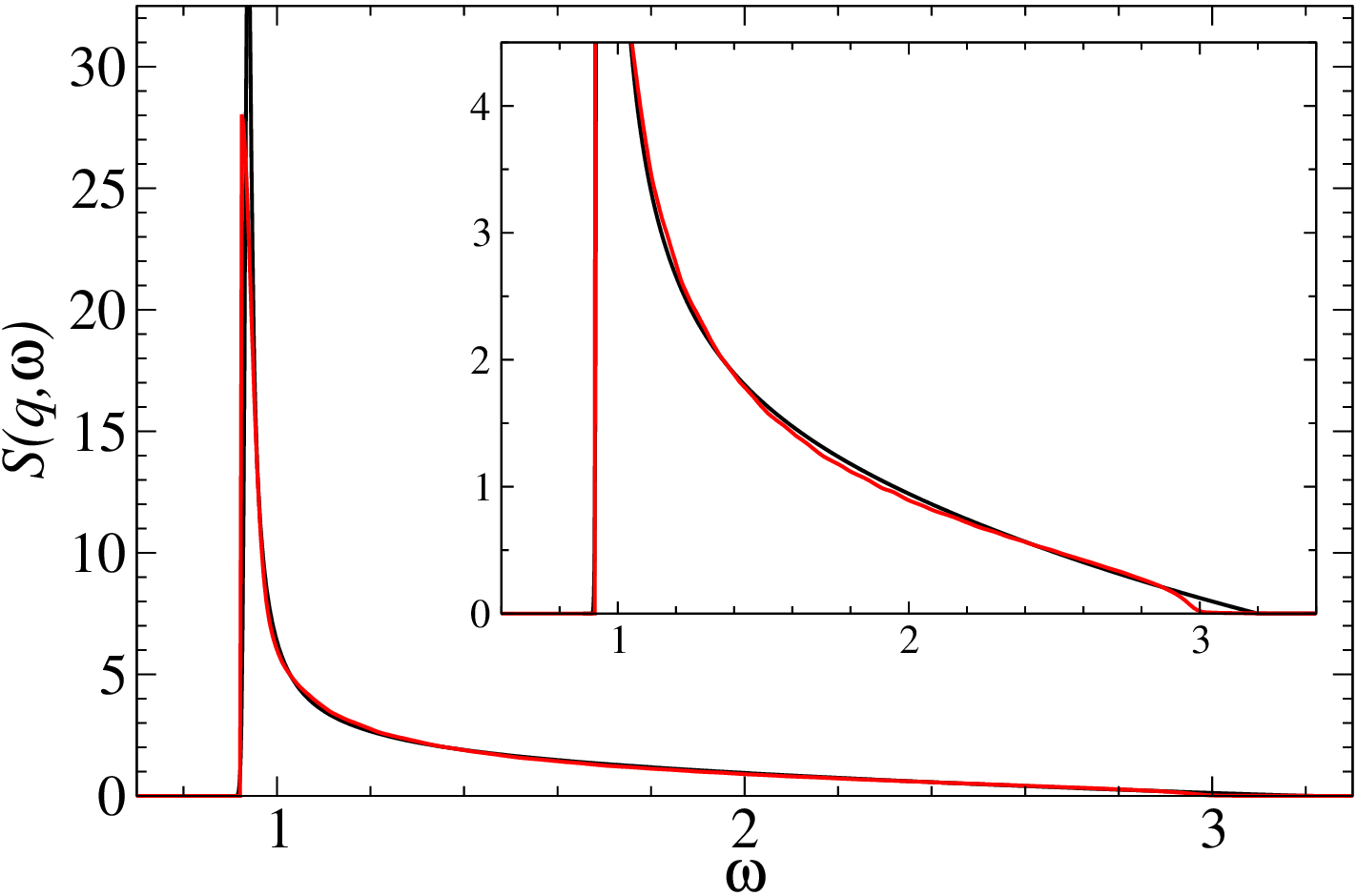}}
\vskip-1mm
\caption{(Color online) $T \to 0$ dynamic structure factor at $q=0.8\pi$ for an $L=500$, obtained after two adjustments of the
spectral bounds (black curve). The BA result \cite{caux05,cauxdata} is shown with the red curve.}
\label{fig3}
\vskip-2mm
\end{figure}

A faster way to identify the spectral bounds is to begin with high upper edge (beyond what is expected for the true spectrum) 
and identify the best lower bound under that condition. With the lower bound fixed at its optimum, the upper bound can be optimized next. Iterating
this procedure once or twice typically leads to excellent bounds very close to those obtained in a two-dimensional search. The results of such 
a procedure for a small spacing, $\Delta_\omega=0.001$, is shown in Fig.~\ref{fig3}. The agreement with the BA calculation (which for $q=0.8\pi$ misses 
about 2\% of the known total spectral weight) is remarkably good, to the author's knowledge unprecedented in QMC studies. The peak location is off by only 
$1\%$, the lower bound slightly below it deviates by less than $0.5\%$ from the true edge, and the non-trivial profile is reproduced. 

{\it Constrained sampling at T$>$0}.---In addition to the entropy-driven leakage of spectral weight outside the correct bounds, there is
another entropic effect in the sampling of the single-peak spectrum at high (physical) temperature. In such a spectrum the volume 
of the accessible configuration space as a function of the peak height $a_m$ (located at the $m$:th $\delta$-function) is given by
\begin{equation}
V(a_m)=\frac{(a_m-a_0)^{m-1}}{(m-1)!}\frac{a_m^{N-m}}{(N-m)!},
\label{volume}
\end{equation}
where $a_0$ is a floor imposed on the spectrum at the low-frequency bound, $a_1 \ge a_0$, which again is regarded as an adjustable parameter 
to be optimized by monitoring $\chi^2(a_0)$. The floor at the high-frequency bound does not appear explicitly, being at $0$ since the spectrum
always decays to $0$ when $\omega \to \infty$, unlike at $w \to 0$. Sampling a spectrum (\ref{aqform}) without any data, i.e., with $\chi^2=0$ in 
Eq.~(\ref{psample1}), the fact that the configurational entropy $\ln(V)$ increases rapidly with $a_m$ will drive the peak to infinite height 
(since no normalization is imposed). Sampling with $\chi^2$ will of course counter-act this effect, but still the entropy will unduly favor a 
sharp peak when $N$ is large. This is not a serious issue in the $T=0$ case discussed above (unless $N$ is much larger than in Fig.~\ref{fig3}), 
because this spectrum has a very sharp peak. However, at high $T$ the peak entropy will cause problems, unless this version of the entropic 
catastrophe is counteracted by dividing the probability (\ref{psample1}) by $V(a_m)$.

In order to obtain continuity as a function of $T$, considering that no entropic counter-weighting was required above at $T=0$, the following probability 
is used
\begin{equation}
P(A) \propto {\rm exp}\bigl(-\chi^2/2-\lambda \ln[V(a_m)]\bigr ),
\label{psample2}
\end{equation}
where $\lambda$ is also to be optimized using $\chi^2(\lambda)$. In practice, it was found that $\lambda=1$ gives good 
solutions when the floor $a_0>0$, while optimizing $\lambda \in [0,1]$ is better when $a_0=0$. Optimizing $\lambda$ after identifying the
spectral bounds in the $T=0$ case discussed above gave $\lambda \approx 0$ and no significant change in the spectrum from Fig.~\ref{fig3}. 
The optimal $\lambda$ varies monotonically as $T$ is increased.

The form (\ref{psample2}) and the optimization procedures can be easily generalized to more than one peak.
An even better form of the probability with entropy suppression may possibly be obtained by using $V(a_m)$ at fixed normalization, which, however, is 
a much more complicated function which has not yet been evaluated in closed form.

\begin{figure}
\centerline{\includegraphics[width=7.5cm, clip]{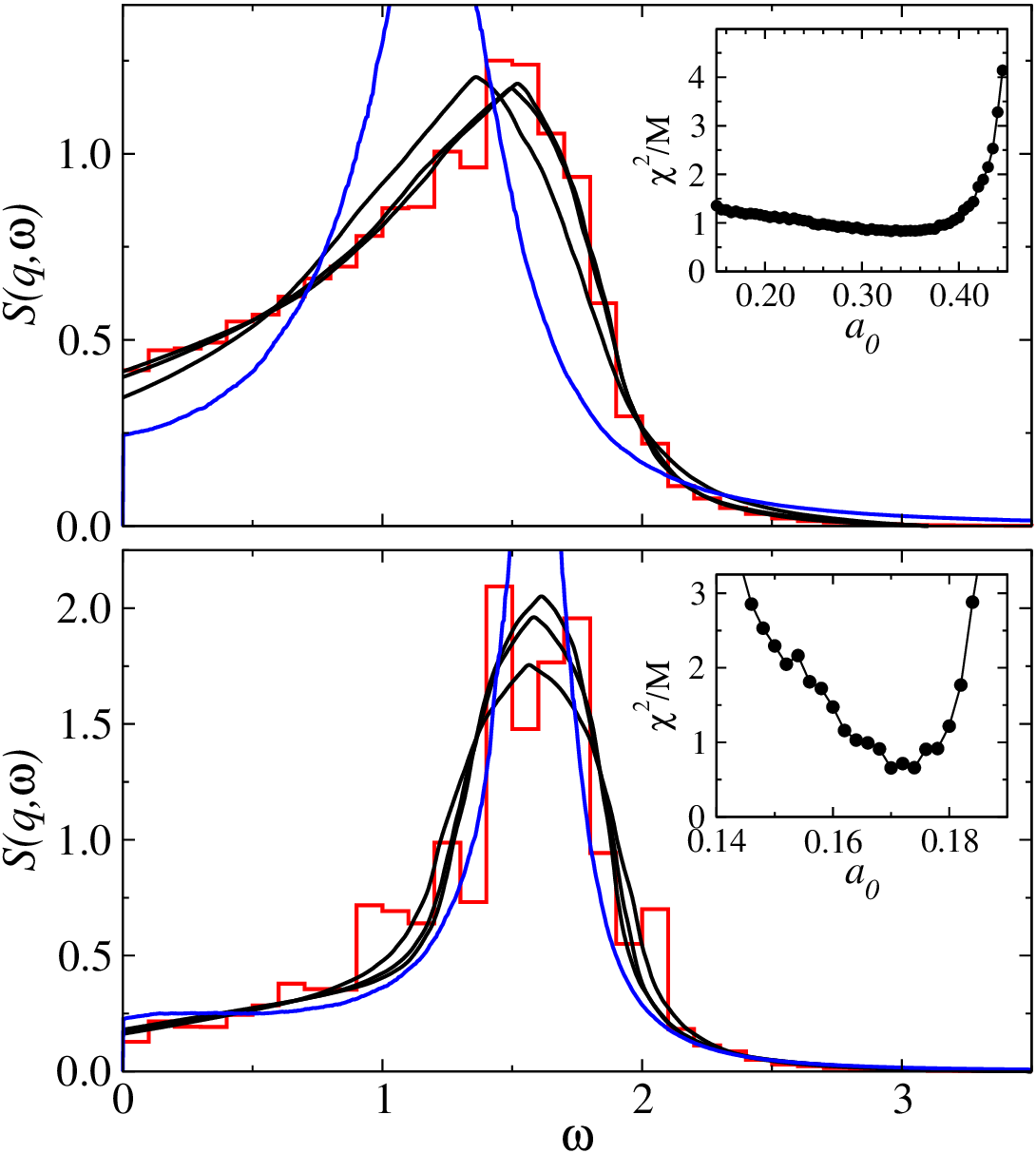}}
\vskip-1mm
\caption{(Color online) Dynamic $q=\pi/2$ structure factor for $L=16$ chains at $T=1$ (top panel) and $0.5$ (bottom panel). The histogram 
(red) represents exact diagonalization results. The black curves were obtained with $\lambda=1$ in Eq.~(\ref{psample2}) and three 
different values of the floor $a_0$; at the minimum $\chi^2(a_0)$ (curves with the lowest $a_0$) and for higher values where $\chi^2/M$ 
is approximately its minimum value plus $M^{-1/2}$ and $2M^{-1/2}$. $\chi^2(a_0)/M$ is shown in the insets.  The curves with higher peaks (blue) 
are from unconstrained sampling. The upper spectral bounds were also chosen according to a $\chi^2$ criterion, as discussed in the text.}
\label{fig4}
\vskip-2mm
\end{figure}

Figure \ref{fig4} shows results at $T=1$ and $1/2$ for an $L=16$ chain, obtained using $\lambda=1$ and scanning over a grid of $a_0$ values.
Exact diagonalization results for the spectrum are represented by 
histograms \cite{starykh97}, and one can of course not expect to resolve the fine structures in such a spectrum by analytic continuation of QMC results.
With the single-peak property imposed one can, however, observe very good agreement with the broad features, including very reasonable values
for the low-energy limit, when choosing $a_0$ such that $\chi^2$ is close to its minimum value. In practice, it is better to go slightly beyond
the floor value minimizing $\chi^2$. When $a_0$ is taken past the minimizing value $\chi^2$ is seen growing rapidly and the spectrum does not 
change much initially in this region, though it changes noticeably at high $T$ for smaller $a_0$. Since the best value $\chi_{\rm min}^2$ 
can fluctuate of the order $M^{1/2}$, it is statistically sound to choose $a_0$ where $\chi^2 \approx \chi^2_{\rm min} + M^{1/2}$, where
the solution typically has stabilized before $\chi^2$ increases sharply. The solution is again not very sensitive to the upper bound as
long as $\omega_N$ is reasonably close to the value to optimizing $\chi^2$. One can determine a suitable bound in an iterative fashion, as discussed 
above, adjusting $a_0$ first with a high $\omega_N$, then adjusting it to where $\chi^2 \approx \chi^2_{\rm min} + M^{1/2}$ (above the point 
where $\chi^2$ is minimized), repeating this once or twice.

Results of this optimized constrained sampling scheme are seen in Fig.~\ref{fig4} to be much better than those of unconstrained sampling, which leads 
to excessively sharp peaks. One can also counteract the peak sharpness in the unconstrained case, e.g., by imposing a ceiling on the weights $a_i$ 
in the sampling. However, results of such a procedure are still not as good as with the constrained sampling, where the form (\ref{volume}) provides 
a more natural mechanism for suppressing the entropy and the shape of the spectrum comes out remarkably well.

{\it Discussion}.---The main result of this work is the identification of configurational entropy as a detriment to stochastic analytic continuation. 
A remarkable improvement in fidelity can be achieved with respect to other methods by suppressing the entropy in various ways. An important aspect of 
these procedures is that the average spectrum no longer depends on the number of $\delta$-functions $N$ used to parametrize it, once $N$ is sufficiently 
large for discretization effects on the scale of the main spectral features to become unimportant.

A bottle-neck of the method is that sampling has to be carried out for many values of the parameters to be optimized; $\omega_1$, $\omega_N$, $a_0$, 
and $\lambda$. However, in practice good results can be obtained with simple scans over a single parameter as follows: For fixed 
$\omega_N$, if $\omega_1=\Delta_\omega/2$ is found to be optimal, then $a_0$ is adjusted with $\lambda=1$. If the optimum is at $a_0=0$, then $\lambda$ 
is optimized. If $\omega_1 > 0$ is optimal one should subsequently also optimize $\lambda$. Very good results for long Heisenberg chains were obtained 
in this way for the full range of temperatures, where comparisons can be made with results of time-dependent density-matrix renormalization 
calculations \cite{barthel09}.

\begin{acknowledgments}
I thank K. Beach, J.-S. Caux, Z. Y. Meng, B. Normand, O. Sylju{\aa}sen, and T. Xiang, for valuable discussions, and
J.-S. Caux also for providing his BA data. This research was supported by the NSF under Grant No.~DMR-1410126 and by 
the Simons Foundation.
\end{acknowledgments}

\end{document}